\documentclass[12pt]{iopart}
\usepackage{iopams,bm}  
\usepackage[dvips]{graphicx}

\newcommand{\diff}{\mathrm{diff}}

\newcommand{\cbar}{\bar{c}}
\newcommand{\Omegabar}{{\bar{\Omega}}}

\newcommand{\rhobar}{\rho^{\mathrm{ref}}}
\newcommand{\pprime}{{p^\prime}}

\newcommand{\Ocal}{{\cal O}}

\newcommand{\bra}[1]{\langle #1|}
\newcommand{\ket}[1]{|#1 \rangle}
\newcommand{\bracket}[2]{\langle #1|#2 \rangle}
\newcommand{\trial}{{\mathrm{(trial)}}}

\newcommand{\rhombus}{
  \put(0,0){\circle*{4}}
  \put(20,0){\circle*{4}}
  \put(10,17.3){\circle*{4}}
  \put(30,17.3){\circle*{4}}
}
\newcommand{\QDMtz}{\unitlength0.05em 
 \begin{minipage}{35\unitlength}
 \begin{center}
 \begin{picture}(30,17)
  \rhombus
  \put(0,0){\line(1,0){20}}
  \put(10,17.3){\line(1,0){20}}
 \end{picture}
 \end{center}
 \end{minipage}
}
\newcommand{\QDMts}{\unitlength0.05em 
 \begin{minipage}{35\unitlength}
 \begin{center}
 \begin{picture}(30,17)
  \rhombus
  \qbezier(0,0)(5,8.65)(10,17.3)
  \qbezier(20,0)(25,8.65)(30,17.3)
 \end{picture}
 \end{center}
 \end{minipage}
}

\begin{document}

\title{Reduced Density Matrices and Topological Order in a Quantum Dimer Model}

\author{Shunsuke Furukawa$^{1,2,3}$, Gr\'egoire Misguich$^2$, Masaki Oshikawa$^4$}
\address{$^1$Laboratoire de Physique Th\'eorique de la Mati\`ere Condens\'ee, UMR 7600 of CNRS, \\
Universit\'e P. et M. Curie, case 121, 4 Place Jussieu, 75252 Paris Cedex, France}
\address{$^2$Service de Physique Th\'eorique, CEA Saclay, 91191 Gif-sur-Yvette Cedex, France}
\address{$^3$Department of Physics, Tokyo Institute of Technology, Meguro-ku, Tokyo 152-8551, Japan}
\address{$^4$Institute for Solid State Physics, University of Tokyo, 5-1-5 Kashiwanoha, Kashiwa 277-8581, Japan}

\begin{abstract}
Resonating valence bond (RVB) liquids in two dimensions are believed 
to exhibit topological order 
and to admit no local order parameter of any kind.
This is a defining property of "liquids"
but it has been explicitly confirmed only in a few exactly solvable models.
In this paper, we investigate the quantum dimer model on the triangular lattice.
It possesses an RVB-type liquid phase, however, for which the absence of 
a local order parameter has not been proved. We examine the question numerically
with a measure based on reduced density matrices. We find a scaling of the
measure which strongly supports the absence of any local order parameter.
\end{abstract}

\pacs{75.10.Jm, 75.50.Ee}
\submitto{\JPCM}

\section{Introduction}
The existence of short-ranged resonating valence bond (RVB) liquids 
has been proposed in several two-dimensional quantum spin models \cite{ml05}. 
These liquids exhibit no kind of simple order 
but the ground states (GS) are degenerate 
if the lattice is put on a surface with a non-trivial topology (cylinder, torus, etc.). 
It is (widely) believed that such degeneracy has a purely topological origin 
and is not ascribed to spontaneous symmetry breaking in terms of any local order parameter. 
This topological nature has been rigorously confirmed 
in a few exactly solvable models \cite{if02,fmo06}, 
but proving this property in more general cases is a challenging issue.

In this paper, we address this issue by formulating the problem 
in terms of reduced density matrices (RDM) as we proposed in Ref. \cite{fmo06}.
In this formulation, we examine the RDMs of the degenerate GSs for various subareas of the system 
and search for an operator distinguishing the degenerate GSs, 
which means that it can be used as an order parameter.
We define a convenient measure for this purpose, which is non-zero for an area 
where an order parameter can be defined.
In conventional orders, the measure is non-zero on some finite local area, 
which indicates the existence of an order parameter on that area.
In RVB liquids, in contrast, it is expected that the measure is zero (in the thermodynamic limit) 
on any local area, thereby verifying topological order.

With this formulation, we analyze the quantum dimer model (QDM) on the triangular lattice, 
which is one of the simplest microscopic models realizing RVB liquids \cite{ms01}.  
We numerically show that the dimer liquid in this model 
cannot be characterized by any local order parameter.

\section{Formulation}

\subsection{General Case}
We formulate a method to detect an order parameter starting from a general setting.
Let $q$ be the degeneracy of the GSs 
and $\ket{\Phi_i}~(i=1,\cdots,q)$ be the orthonormal GSs. 
For a state $\ket{\Psi}$ in the ground-state subspace, 
we define its RDM on an area $\Omega$ 
by tracing out the degrees of freedom outside $\Omega$:
$\rho_\Omega \equiv \Tr_\Omegabar \ket{\Psi}\bra{\Psi}$, 
where $\Omegabar$ is the complement of $\Omega$. 
We also introduce as a reference the RDM averaged over the ground-state subspace: 
$\rhobar_\Omega \equiv \frac1q \Tr_{\Omegabar} \sum_{i=1}^{q} \ket{\Phi_i}\bra{\Phi_i}$. 
Note that $\rhobar_\Omega$ is independent of the choice of the basis 
$\{\ket{\Phi_i}\}$ of the ground-state subspace.
An order parameter can be defined on $\Omega$ 
if and only if there exists $\ket{\Psi}$ such that $\rho_\Omega \ne \rhobar_\Omega$. 

To quantify to what extent RDMs are distinguishable, 
we introduce a measure $D_\Omega$ as follows. 
We first define a measure of difference between two RDMs as \cite{fmo06}
\begin{equation}\label{eq:diff}
  \diff (\rho_\Omega,\rhobar_\Omega) \equiv \max_{|\Ocal_\Omega|\le 1} 
  \bigg|\Tr_{\Omega} (\Ocal_\Omega\rho_\Omega)
       -\Tr_{\Omega} (\Ocal_\Omega\rhobar_\Omega)\bigg|, 
\end{equation}
where $\Ocal_\Omega$ is a (variational) Hermitian operator on $\Omega$ 
whose norm is less than unity, i.e., 
$|\bra{\psi}\Ocal_\Omega\ket{\psi}| \le 1$ for any normalized vector $\ket{\psi}$. 
Using the eigenvalues $\{\lambda_j\}$ of $\rho_\Omega-\rhobar_\Omega$, 
this can be simplified as: 
$\diff(\rho_\Omega,\rhobar_\Omega)=\sum_j |\lambda_j|$.
We define $D_\Omega$ by maximizing $\diff(\rho_\Omega,\rhobar_\Omega)$ 
over the choice of a GS $\ket{\Psi}$: 
\begin{equation}\label{eq:domega}
  D_\Omega \equiv \max_{\ket{\Psi}} \diff(\rho_\Omega,\rhobar_\Omega).
\end{equation}
The value of $D_\Omega$ in the thermodynamic limit tells us 
whether an order parameter can be defined on $\Omega$. 
The measure $D_\Omega$ has the following useful properties: 
(a) Normalization to a definite range $0\le D_\Omega \le 2-2/q$. 
(b) Monotonicity: if an area $\Lambda$ completely contains an area $\Omega$, 
we have $D_\Omega\le D_\Lambda$.  

How to calculate $D_\Omega$ is not clear from the definition. 
In the case of QDMs, a simple formula is available, 
which we derive in the next subsection.
However, for the completeness of the method, 
here we briefly describe a general algorithm called {\it iterative maximization}.
The difficulty in calculating $D_\Omega$ resides in the double maximizations 
in Eqs. \ref{eq:diff} and \ref{eq:domega}, 
but we notice that each maximization is possible if either $\Ocal_\Omega$ or $\ket{\Psi}$ is fixed.
The idea is to alternately perform these two kinds of maximizations starting from 
a random vector $\ket{\Psi}$.
This would lead at least to a local maximum in the space of $\Ocal_\Omega$ and $\ket{\Psi}$.
Starting from various random vectors, we would obtain the global maximum $D_\Omega$.
This method has a potential to handle a wide range of problems.

\subsection{Case of Quantum Dimer Models}

We limit the setting to the case of gapped dimer liquids in QDMs on two-dimensional lattices, 
where further argument is possible. 
The gapped dimer liquids have been found in several QDMs \cite{ms01,msp02} 
and related models \cite{if02,k03}. 
However, except for the special cases \cite{if02,fmo06,msp02} 
where the correlation length is strictly zero, 
it is difficult to prove the absence of all the local order parameter. 
This motivates us to study more general cases.

Let $S$ be the set of all the dimer coverings of the lattice.
If the lattice has a non-trivial topology, 
$S$ can be grouped into topological sectors 
which are not mixed by any local operation.
We concentrate on the case of the torus in the following.
We draw two incontractible loops $\Delta_1$ and $\Delta_2$ 
which pass through the bonds and wind the torus in $x$ and $y$ directions respectively. 
We classify $S$ into four topological sectors $S^p$ with $p=++,+-,-+,--$, 
depending on the parity of the number of dimers crossing $\Delta_1$ and $\Delta_2$.
In a gapped dimer liquid, 
the lowest-energy states $\ket{\Phi_p}$ in different topological sectors $S^p$ 
become degenerate in the thermodynamic limit.
We employ $\ket{\Phi_p}$'s as the basis of the ground-state subspace.

To define RDMs for QDMs, we must specify the local degrees of freedom of the models.
To this end, we assign an Ising variable $\sigma_k$ to each bond $k$ of the lattice 
as in Ref. \cite{msf02} 
and identify the presence/absence of a dimer on the bond as $\sigma_k=+1$ and $-1$, respectively. 
Any physical configuration $\{\sigma_k\}$ must satisfy the hard-core constraints: 
for each site of the lattice, there must be exactly one bond with $\sigma_k=1$ emanating from it. 
An area $\Omega$ is defined as a set of bonds. 
We define the matrix element of the RDM of a GS $\ket{\Psi}$ as
\begin{equation}
 \bra{c_1}\rho_\Omega\ket{c_2} 
 = \sum_{\cbar} \bracket{c_1,\cbar}{\Psi} \bracket{\Psi}{c_2,\cbar}, 
\end{equation}
where $c_1$ and $c_2$ are dimer configurations on $\Omega$ 
and the sum is over all the dimer configurations $\cbar$ on $\Omegabar$.
Note that we set $\bracket{c,\cbar}{\Psi}=0$ if $(c,\cbar)$ is an unphysical configuration
(violating the hard-core constraint).

We proceed to the evaluation of $D_\Omega$.  
We first assume that the area $\Omega$ is topologically non-trivial, i.e., 
$\Omega$ encircles the torus.
Without loss of generality, we assume that $\Omega$ contains $\Delta_1$.
Let us consider the following trial GS and trial operator:
\[
 \ket{\Psi^\trial}=\ket{\Phi_{++}}, \qquad 
 \Ocal_\Omega^\trial=\sum_c \ket{c} P(c) \bra{c}, 
\]
where in the second equation $c$ runs over all the dimer configurations on $\Omega$ 
and $P(c)$ is its parity along $\Delta_1$. 
These gives a lower bound of $D_\Omega$:
\[
  D_\Omega \ge \bigg| \Tr_\Omega \Ocal_\Omega^\trial 
                      \left( \Tr_\Omegabar \ket{\Psi^\trial}\bra{\Psi^\trial} - \rhobar_\Omega \right)\bigg|
           = 1
\]
Thus the value of $D_\Omega$ is consistent with the existence of a non-local order parameter.

We next assume that $\Omega$ is a (finite) local area.
In this case, we can prove the following relation:
\begin{equation}\label{eq:qdmlocal}
 \Tr_\Omegabar \ket{\Phi_p}\bra{\Phi_\pprime} =0 \quad (p\ne\pprime ).
\end{equation}
To prove this relation, we explicitly express the matrix element of the LHS: 
\[
 \bra{c_1} \bigg( \Tr_\Omegabar \ket{\Phi_p}\bra{\Phi_\pprime} \bigg) \ket{c_2}
 = \sum_{\cbar} \bracket{c_1,\cbar}{\Phi_p} \bracket{\Phi_\pprime}{c_2,\cbar}
\]
Since $\Omega$ is local, we can choose $\Delta_1$ and $\Delta_2$ so as not to touch $\Omega$.
Then the dimer coverings $(c_1,\cbar)$ and $(c_2,\cbar)$ always belong to a common topological sector 
and therefore $\bracket{c_1,\cbar}{\Phi_p}$ and $\bracket{\Phi_\pprime}{c_2,\cbar}$ cannot become non-zero 
at the same time.
Hence we obtain Eq. \ref{eq:qdmlocal}.
Using Eq. \ref{eq:qdmlocal}, we can derive a simpler expression for $D_\Omega$: 
\begin{equation}
 D_\Omega = \max_p \diff(\rho_\Omega^p,\rhobar_\Omega), 
  \quad\mathrm{with}\enskip 
 \rho_\Omega^p \equiv \Tr_\Omegabar \ket{\Phi_p}\bra{\Phi_p}. 
\end{equation}
This is computable numerically and is employed in the calculation shown below.

\section{Numerical Result}
We consider the QDM on the triangular lattice. The Hamiltonian reads \cite{ms01}: 
\begin{equation}
\hspace{-2cm}
H=\sum_{\mathrm{rhombi}}\left[-t\left(\bigg|\QDMts\bigg\rangle
                 \bigg\langle\QDMtz\bigg|+\mathrm{h.c.}\right)
            +v\left(\bigg|\QDMtz\bigg\rangle\bigg\langle\QDMtz\bigg|
     +\bigg|\QDMts\bigg\rangle\bigg\langle\QDMts\bigg|\right)\right].
\end{equation}
At the Rokhsar-Kivelson (RK) point ($v=t$), the GS in each sector is exactly 
the equal-amplitude superposition of all the dimer coverings belonging to that sector \cite{rk88,ms01}:
\begin{equation}\label{eq:RK-fn}
 \ket{\Phi_p} = \frac{1}{\sqrt{|S^p|}} \sum_{C\in S^p} \ket{C}.
\end{equation}
It has been numerically shown that the degeneracy of the GSs 
and the exponential decay of the dimer-dimer correlation at the RK point 
persist up to some range in the parameter space, 
forming a dimer liquid phase in $0.7 \lesssim v/t \le 1$ \cite{ms01,rfbim05,ifiitb02,iif02}.

\begin{figure}[h]
 \begin{center}
 \begin{picture}(200,140)
  \put(-30,-30){\includegraphics[width=9cm,clip]{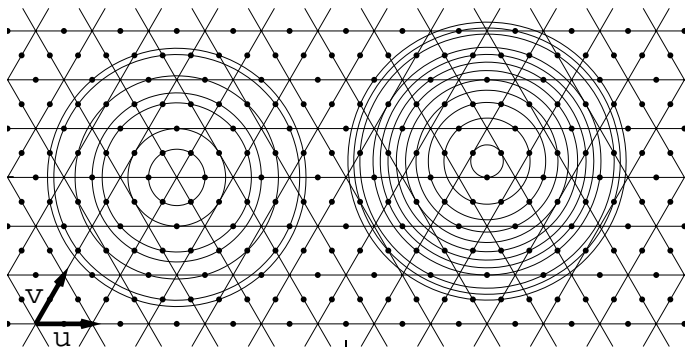}}
 \end{picture}
 \end{center}
\caption{Circular areas centered around a site (left) or an interior of a triangle (right). }
\label{fig:circular-areas}
\end{figure}

\begin{figure}[h]
 \begin{center}
  \includegraphics[width=7cm,clip]{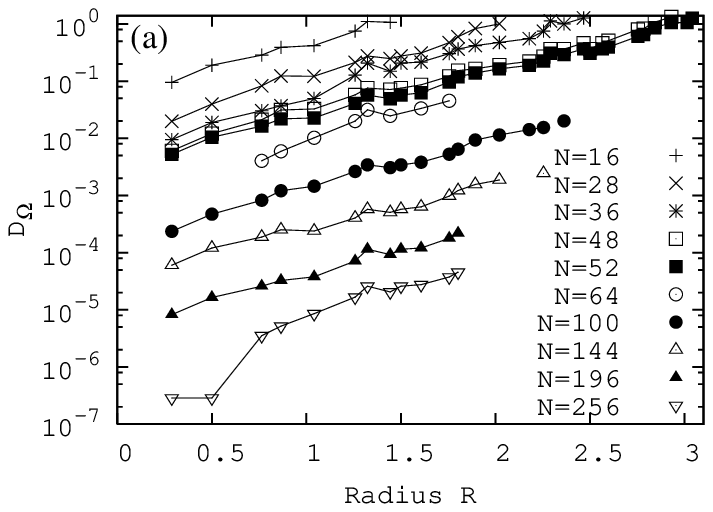}
  \includegraphics[width=7cm,clip]{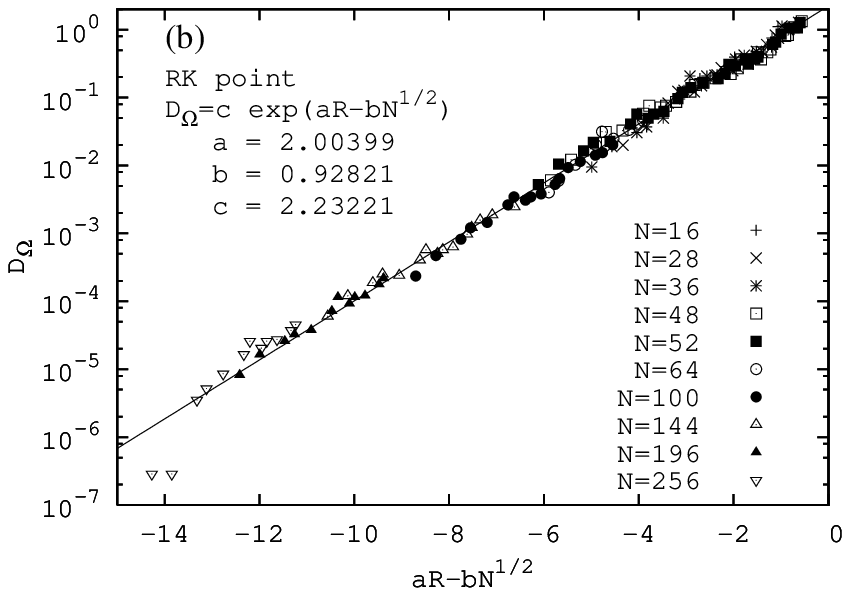}
 \end{center}
\caption{The result for the RK wave function. 
(a) The value of $D_\Omega$ as a function of the radius $R$ 
for different system sizes $N$. 
(b) Fitting of the data using an exponential function $D_\Omega\approx c e^{aR-b\sqrt{N}}$.}
\label{fig:diff-RK}
\end{figure}

\begin{figure}[h]
 \begin{center}
  \includegraphics[width=7cm,clip]{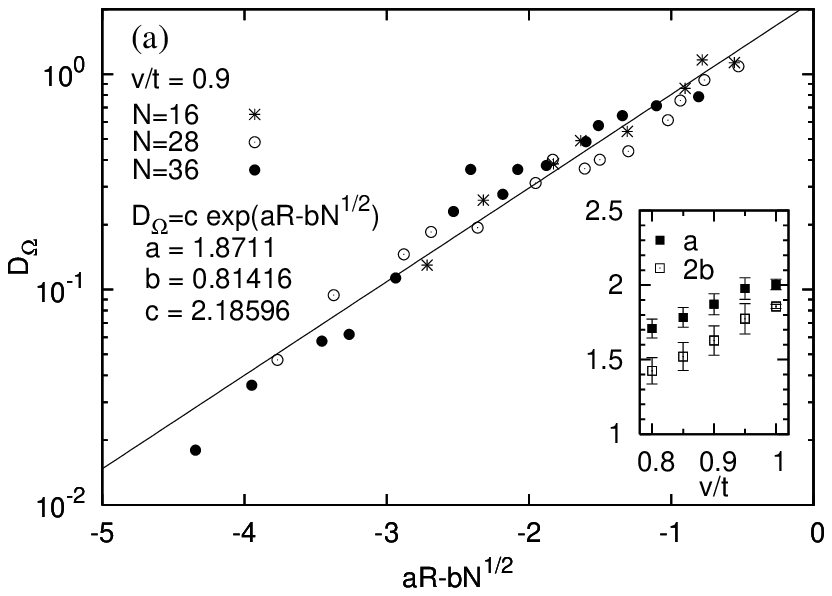}
  \includegraphics[width=7cm,clip]{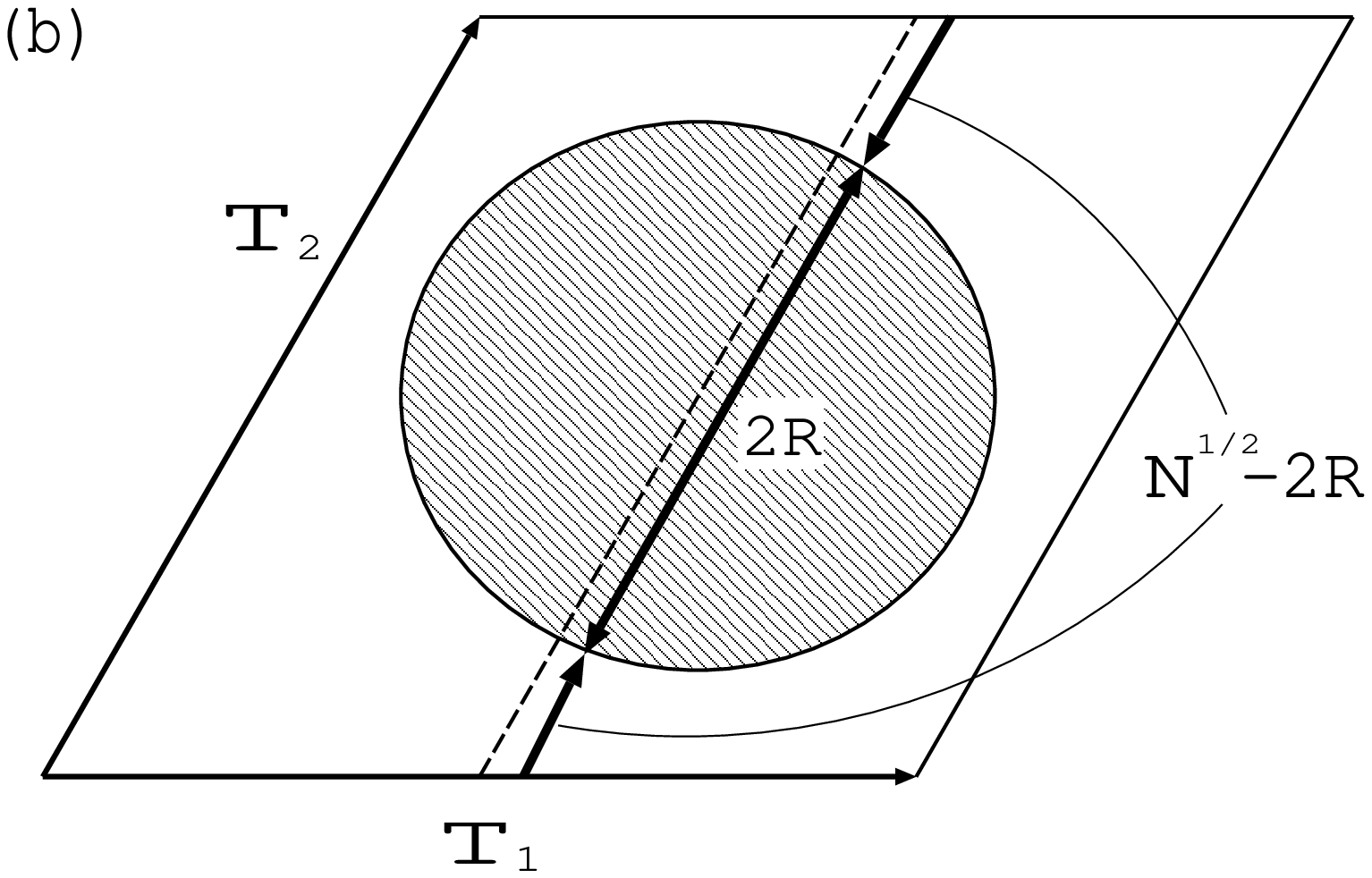}
 \end{center}
\caption{(a) Main panel: fitting of the data of $D_\Omega$ for $v/t=0.9$. 
Inset: obtained fitting parameters $a,b$ for different values of $v/t$ in the dimer liquid phase.
(b) A loop (broken line) encircling the torus and the area $\Omega$. 
Periodic boundary conditions are imposed along $\bm{T}_1$ and $\bm{T}_2$. 
A part in the loop with the length of $\sqrt{N}-2R$ is not covered by $\Omega$.}
\label{fig:diff-non-RK}
\end{figure}

We first numerically calculated $D_\Omega$ for the RK wave function, Eq. \ref{eq:RK-fn}, 
through the direct enumeration of dimer coverings (up to $N=52$)
or the evaluation of Pfaffians of Kasteleyn matrices \cite{k61} (up to $N=256$).
The former method is suited for large areas while the latter can handle large systems.
The lattice is put on the torus and is defined by two vectors $\bm{T}_1$ and $\bm{T}_2$
specifying the periodicity.
We require that the lattice be symmetric under the $2\pi/3$ rotation, choosing 
$\bm{T}_1=l\bm{u}+m\bm{v},\bm{T}_2=-m\bm{u}+(l+m)\bm{v}$, 
where $l$ and $m$ are integers 
and $\bm{u}$ and $\bm{v}$ are unit vectors as shown in Fig. \ref{fig:circular-areas}.
The total number of sites is given by $N=l^2+lm+m^2$.
As the choice of $\Omega$, we define circular areas in the following way: 
we draw a circle centered around a site or an interior of a triangle 
and regard every bond whose midpoint is in the circle as an element of the area 
(see Fig. \ref{fig:circular-areas}). 
The values of $D_\Omega$ for circular areas are plotted versus the radius $R$ 
in Fig. \ref{fig:diff-RK} (a).  
Though $D_\Omega$ tends to increase as a function of $R$, 
it remains small if $R$ is sufficiently smaller than the linear system size $\sqrt{N}$.
It appears that an exponential dependence 
$D_\Omega\approx c e^{aR-b\sqrt{N}}$ fits the data relatively well,  
as can be seen in Fig. \ref{fig:diff-RK} (b).
We also performed calculation for $v/t<1$ 
by obtaining GSs by exact diagonalization.
We fit the data in the same way and 
the resultant parameters $a, b$ are shown in the inset of Fig. \ref{fig:diff-non-RK} (a).
We observe that the relation $a\approx 2b$ holds approximately in the dimer liquid phase.
This indicates that the finite size effect in $D_\Omega$ enters through $\sqrt{N}-2R$.
To interpret this, let us consider a loop encircling the torus and passing through $\Omega$; 
see Fig. \ref{fig:diff-non-RK} (b).  
Some part of a loop is not covered by $\Omega$ 
and the minimal length of the part is $\sqrt{N}-2R$.
The lack of this part makes the distinction of sectors obscure, 
hence the exponential reduction in $D_\Omega$.
Our result explicitly shows the topological nature of the liquid
since $D_\Omega$ goes to zero exponentially with $N\to\infty$ {\em for any fixed $R$}.
This does not only exclude local order parameters which are diagonal in the dimer basis 
(and amenable to exact calculations in the $N\to\infty$ limit \cite{iif02}), 
but any, possibly non-diagonal, operator.
The parameter $b$ is naturally interpreted as the inverse of a correlation length, 
and the value we obtained at the RK point is close to  
$\xi_A^{-1}\approx 0.83\pm0.12i\pi$ and $\xi_B^{-1}\approx 0.76$ 
obtained by Ref. \cite{iif02}, 
where $A$ refers to the direction along a bond and $B$ the direction perpendicular to it.
 

\section{Conclusion}
We formulated a method to detect the existence of an order parameter 
using reduced density matrices.
We numerically applied the method to the liquid phase of the QDM on the triangular lattice. 
The measure $D_\Omega$ we defined contrasts local areas with non-local areas:
$D_\Omega\to 0$ with $N\to \infty$ on local areas while $D_\Omega\ge 1$ on non-local areas. 
The data of $D_\Omega$ is fitted well by an exponential function, 
which can be interpreted from the topological picture.
Our result explicitly verifies topological order in this system.

\bigskip
We are grateful to C. Lhuillier and V. Pasquier for many fruitful discussions.
SF and MO were supported by a 21st Century COE Program at Tokyo Tech, 
"Nanometer-Scale Quantum Physics", and 
SF was also supported by an advanced student exchange pilot project, 
"Coll\`ege Doctoral Franco-Japonais", both from the MEXT of Japan.

\end{document}